\documentclass{article}
\usepackage{spconf,amsmath,graphicx}

\usepackage{cite}
\usepackage{amssymb,amsfonts}
\usepackage{algorithmic}
\usepackage{textcomp}
\usepackage{xcolor}
\usepackage{float}
\usepackage{amsthm}
\usepackage{graphicx}
\usepackage{epstopdf}
\usepackage{amsmath,bm}
\usepackage{amsfonts}
\usepackage{amssymb}
\usepackage{color}
\usepackage{multirow}
\usepackage{multicol}
\usepackage{soul,xcolor}

\definecolor{orange}{RGB}{0,112,192}
\DeclareMathOperator*{\argmin}{arg\,min}

\title{ADMM-BASED ONE-BIT QUANTIZED SIGNAL DETECTION FOR MASSIVE MIMO SYSTEMS WITH HARDWARE IMPAIRMENTS}
\name{\"Ozlem Tugfe Demir and Emil Bj\"ornson \thanks{This work was partially supported by ELLIIT and the Wallenberg AI, Autonomous Systems and Software Program (WASP) funded by the Knut and Alice Wallenberg Foundation.}}
\address{{Department of Electrical Engineering (ISY), Link\"oping University, Sweden
	} \\
	{Email: \{ozlem.tugfe.demir, emil.bjornson\}@liu.se}}

\begin{document}
\ninept
\maketitle
\begin{abstract}
This paper considers signal detection in massive multiple-input multiple-output (MIMO) systems with general additive hardware impairments and one-bit quantization. First, we present the quantiza-tion-unaware and Bussgang decomposition-based linear receivers by generalizing them for the considered hardware impairment model. We propose an optimization problem to estimate the uplink data signals by choosing a suitable cost function that treats the unquantized received signal at the base station as the variable. We exploit the additional structure of the one-bit quantization and signal modulation by including proper constraints. To solve the non-convex quadratically-constrained quadratic programming (QCQP) problem, we propose an ADMM-based algorithm with closed-form update equations. Then, we replace the harsh projectors in the updates with their soft versions to improve the detection performance. We show that the proposed ADMM-based algorithm outperforms the state-of-the-art linear receivers significantly in terms of bit error rate (BER) and the performance gain increases with the number of antennas and users.
\end{abstract}
\vspace{-0.2cm}
\begin{keywords}
MIMO signal detection, massive MIMO, one-bit quantization, ADMM, hardware impairments.
\end{keywords}
\vspace{-0.3cm}
\section{Introduction}
 \vspace{-0.2cm}
\label{sec:intro}
Massive MIMO (multiple-input multiple-output) is one of the key components of 5G cellular systems and commercial deployments began in 2018 \cite{massive_mimo_reality}. Deploying a large number of antennas at the base station (BS) to support multiple users on the same time-frequency channel is one of the main ingredients of this technology \cite{one_bit_massive_mimo}. Several practical concerns arise in implementing massive MIMO technology compared to the previous cellular systems equipped with moderate numbers of antennas. Using near-distortionless hardware is highly costly for very large number of antennas \cite{near_ml_detector}. Some papers in the literature studied the effect of hardware impairments such as phase errors, low-noise amplifier non-linearities, or stochastic additive noise on massive MIMO  \cite{emil_nonideal,emil_book,emil_hardware,bruno_hardware,ozlem_iswcs}. In addition, several works considered low-cost analog-to-digital converters (ADC) with one-bit quantization \cite{one_bit_massive_mimo,near_ml_detector,one_bit_massive_mimo2,one_bit_receivers,ber_floor}. However, most of the existing works which analyze one-bit quantized massive MIMO have not considered other distortions caused by amplifiers, local oscillators, and mixers. To the best of the authors' knowledge, this paper is the first work which considers uplink signal detection in massive MIMO under general additive distortion noise together with one-bit quantization.
 \vspace{-0.2cm}
\subsection{Related Works}
\vspace{-0.1cm}
 Signal estimation and detection from one-bit quantized samples have been studied outside the massive MIMO literature \cite{signal_recovery,deep_one_bit, one_bit_new1, one_bit_new2, one_bit_new3, one_bit_new4}. \cite{signal_recovery} considered estimation of an unknown single parameter in a wireless sensor network with multiple single-antenna nodes.  On the other hand, \cite{deep_one_bit} used the deep unfolding technique to solve multi-dimensional signal recovery problem with one-bit quantization.

In this paper, we use the idea from \cite{signal_recovery} of casting the signal detection as an optimization problem to solve a multi-user QPSK detection by including modulation constraints. We reformulate the original non-convex problem to solve it with the powerful  Alternating Direction Method of Multipliers (ADMM) algorithm \cite{boyd_admm} and we obtain closed-form updates. Recently, ADMM has been used for signal detection   for a perfect large-scale MIMO transceiver system in \cite{lasso-admm}.
\subsection{Contributions}
\begin{itemize}
	\item To the best of authors' knowledge, this paper is the first work which considers the joint effect of the stochastic additive hardware impairment model considered in \cite{emil_book,emil_nonideal,bruno_hardware} and  one-bit ADCs on the uplink signal detection. 
	\item Unlike the previous work that considers one-bit quantized massive MIMO, in our model, we have a conditional colored Gaussian distortion. We generalize the Bussgang decomposition-based  linear receivers proposed recently in \cite{one_bit_receivers} by considering the joint hardware impairment model.

\item In order to exploit the transmitted signal characteristics, we cast a new optimization problem for QPSK signaling and reformulate it to achieve an ADMM-based algorithm with closed-form update equations.
\item We propose a softening method by replacing the harsh-nature functions in the updates with their soft versions to improve the detection performance. 
\end{itemize}
We verify the effectiveness of the proposed algorithm for several scenarios and show that the ADMM-based algorithm outperforms the conventional linear receivers substantially.

\section{System Model}
We consider a single-cell massive MIMO system where a BS equipped with $M$ antennas serves $K$ single-antenna user equipments (UEs). We focus on the uplink with non-ideal BS and UE hardware. A block-fading model is considered where the wireless channels between each BS antenna and UE is represented by a constant complex-valued scalar that takes an independent realization in each time-frequency coherence block \cite{emil_book}. 

 Let ${\bf g}_k=[ \ g_{k1} \ \ldots \ g_{kM} \ ]^T \in \mathbb{C}^{M}$ denote the channel for the $k^{\textrm{th}}$ UE in vector form where $g_{km} \in \mathbb{C}$ is the channel between the $k^{\textrm{th}}$ UE and the $m^{\textrm{th}}$ antenna of the BS. In this paper, we assume that the channels are known at the BS to study the joint effect of hardware impairments and one-bit quantization on the uplink signal detection. Investigating channel estimation is left as future work.

Let $s_k \in  \mathbb{C}$ denote the unit-power information symbol transmitted by the $k^{\textrm{th}}$ UE and $p_k$ is the corresponding transmission power. Based on the established model in \cite{emil_book,emil_nonideal}, the received signal at the BS, ${\bf y}\in \mathbb{C}^{M}$ under both BS and UE non-linear hardware impairments is
\begin{align}\label{eq:received}
{\bf y}=\sqrt{\kappa^r}\sum_{k=1}^K{\bf g}_k\left(\sqrt{\kappa_k^tp_k}s_k+ \eta_k^t\right)+\bm{\eta^r}+{\bf n},
\end{align}
where $\eta_{k}^t \sim \mathcal{N}_{\mathbb{C}}(0,(1-\kappa_k^t)p_k)$ is the distortion noise caused by the $k^{\textrm{th}}$ UE's (transmitter) hardware with the corresponding quality coefficient $\kappa^t_k\in(0,1]$. When $\kappa^t_k=1$, there is no distortion noise and as $\kappa^t_k$ decreases, the distortion noise variance increases. In a similar manner, $\bm{\eta^r} =[ \ \eta^r_1 \ \ldots \eta^r_M \ ]^T \in \mathbb{C}^{M}$ models the hardware distortion at the BS with hardware quality coefficient $\kappa^r\in (0,1]$. $\bm{\eta^r}$ is conditionally Gaussian given the channels $\{{\bf g}_k\}$, i.e. $\eta^r_m|\{g_{km}\}\sim \mathcal{N}_{\mathbb{C}}(0,(1-\kappa^r)\sum_{k=1}^Kp_k|g_{km}|^2)$. ${\bf n} \sim \mathcal{N}_{\mathbb{C}}({\bf 0}_M, \sigma^2{\bf I}_M)$ is the additive white Gaussian noise. Note that the distortions due to the different UEs and BS antenna hardware are independent. In fact, \cite{emil_hardware} showed that the distortion correlation between different BS antennas can be neglected for hardware non-linearities in the uplink as long as the number of UEs is sufficiently large ($>$5), which is of main interest for massive MIMO systems. However, for one-bit quantization distortions, this is not the case and hence, we consider them separately. 

Let us define $\bm{\mu}\triangleq\sqrt{\kappa^r}\sum_{k=1}^K{\bf g}_k\eta_{k}^t+\bm{\eta}^r+{\bf n}\in \mathbb{C}^{M}$ which is the effective colored Gaussian noise given the channel vectors $\{{\bf g}_k\}$ with the conditional distribution $\bm{\mu} |\{{\bf g}_k\} \sim \mathcal{N}_{\mathbb{C}}({\bf 0}_M,{\bf \Sigma})$ where the covariance matrix ${\bf \Sigma} \in \mathbb{C}^{M \times M}$ is given by
\begin{align} \label{eq:Sigma}
{\bf \Sigma}\triangleq \kappa^r\sum_{k=1}^K\left(1-\kappa_k^t\right)p_k{\bf g}_k{\bf g}_k^H+{\bf D},
\end{align}
where ${\bf D} \in \mathbb{C}^{M \times M}$ is a diagonal matrix with the $(m,m)$th entry being $(1-\kappa^r)\sum_{k=1}^Kp_k|g_{km}|^2+\sigma^2$. Note that the effective noise becomes colored with non-diagonal covariance matrix under hardware impairments. ${\bf \Sigma}$ is dependent on the channel realizations and $\bm{\mu}$ is conditionally Gaussian given $\{{\bf g}_k\}$.

Now, we express the quantized signal at the BS by using one-bit ADCs for the real and imaginary parts of the received signal  in \eqref{eq:received}. Let us define the effective channel ${\bf \tilde{g}}_k\triangleq\sqrt{\kappa^r\kappa_k^tp_k}{\bf g}_k\in \mathbb{C}^{M}$, the concatenated channel matrix ${\bf \tilde{G}}\triangleq [ \ {\bf \tilde{g}}_1 \ \ldots \ {\bf \tilde{g}}_K \ ] \in \mathbb{C}^{M \times K}$, and the data vector ${\bf s}\triangleq [ \ s_1 \ \ldots \ s_K \ ]^T\in \mathbb{C}^{K}$. Using these definitions, the received signal in \eqref{eq:received} before quantization can be expressed as ${\bf y}={\bf \tilde{G}}{\bf s}+\bm{\mu}$. Note that the real and imaginary parts of the elements of ${\bf y}$ are one-bit quantized separately, hence it is useful to express the one-bit ADC operation in terms of real variables. The received signal at the BS before quantization can be expressed as ${\bf z}={\bf H}{\bf x}+{\bf v}$ in terms of the real variables
\begin{align} 
&{\bf z}=\begin{bmatrix} \Re\{{\bf y}\} \\ \Im\{{\bf y}\} \end{bmatrix} \in \mathbb{R}^{2M}, \ \ \  {\bf x}=\begin{bmatrix} \Re\{{\bf s}\} \\ \Im\{{\bf s}\} \end{bmatrix} \in \mathbb{R}^{2K} \label{eq:z-x}, \ \ \ \\
&{\bf H}=\begin{bmatrix} \Re\{{\bf \tilde{G}}\} & -\Im\{{\bf \tilde{G}}\} \\ \Im\{{\bf \tilde{G}}\} & \Re\{{\bf \tilde{G}}\}\end{bmatrix} \in \mathbb{R}^{2M \times 2K}, \ \ \ {\bf v}=\begin{bmatrix} \Re\{\bm{\mu}\} \\ \Im\{\bm{\mu}\} \end{bmatrix} \in \mathbb{R}^{2M} \label{eq:H-v}.
\end{align}
Note that the effective noise $\bm{\mu}$ is conditionally circulary symmetric given ${\bf \tilde{G}}$, and hence the covariance matrix of the real noise vector ${\bf v}$ in \eqref{eq:H-v} is given by
\begin{align} \label{eq:C}
{\bf C}=\frac{1}{2}\begin{bmatrix} \Re\{{\bf \Sigma}\} & -\Im\{{\bf \Sigma}\} \\ \Im\{{\bf \Sigma}\} & \Re\{{\bf \Sigma}\} \end{bmatrix}\in \mathbb{R}^{2M \times 2M}.
\end{align}
Then, the received signal after one-bit quantization is given by ${\bf r}=\text{sgn}({\bf z})$ where $\text{sgn}({\bf z})$ is the sign function which is applied to the elements of the vector ${\bf z}$ individually, i.e., $r_m=1$ if $z_m\geq 0$ and $r_m=-1$ otherwise, for $m=1,\ldots,2M$. 

We present possible linear receivers for the quantized and impaired massive MIMO system in the following section.

\section{Linear Receivers}
We now present the quantization-unaware and Bussgang-based quantization-aware linear receivers from \cite{one_bit_receivers}. Note that the quan-tization-aware receivers in \cite{one_bit_receivers} are proposed only for massive MIMO systems with the perfect transceivers except for one-bit quantization. In this section, we generalize them by taking the joint effect of hardware impairments and one-bit quantization into account.

Let ${\bf W}=[ \ {\bf w}_1 \ \ldots \ {\bf w}_K \ ]^T\in \mathbb{C}^{K\times M}$ be the receive combining matrix and the signal for the $k^{th}$ user's data detection is given by ${\bf w}_k^T{\bf \tilde{r}}$, for $k=1,\ldots,K$, where the complex quantized signal ${\bf \tilde{r}} \in \mathbb{C}^{M}$ is given by
\begin{align} \label{eq:tilder}
{\bf \tilde{r}}=\left(1/\sqrt{2}\right)\text{sgn}(\Re\{{\bf y}\})+\left(j/\sqrt{2}\right)\text{sgn}(\Im\{{\bf y}\}).
\end{align}

\subsection{Quantization-Unaware Linear Receivers}
The quantization-unaware receivers simply neglect the effect of one-bit quantization, however, they take into account the hardware impairments in \eqref{eq:received}. The conventional receivers in massive MIMO are maximum ratio combining (MRC), zero-forcing (ZF), and minimum mean-squared error (MMSE) receivers \cite{erik_book,emil_book} which are given by
\begin{align}
& {\bf W}_{\text{MRC}}={\bf \tilde{G}}^H, \label{eq:qun_mrc} \\
& {\bf W}_{\text{ZF}}=({\bf \tilde{G}}^H{\bf \tilde{G}})^{-1}{\bf \tilde{G}}^H, \label{eq:qun_zf} \\
&{\bf W}_{\text{MMSE}}={\bf \tilde{G}}^H({\bf \tilde{G}}{\bf \tilde{G}}^H+{\bf \Sigma})^{-1}. \label{eq:qun_mmse}
\end{align}
Note that they neglect the one-bit quantization effect and MMSE receiver treats the distortion $\bm{\mu}$ as colored noise. 

\subsection{Quantization-Aware Linear Receivers} 
Let us use the Bussgang decomposition \cite{one_bit_receivers} to express the one-bit quantized complex signal in \eqref{eq:tilder} as ${\bf \tilde{r}}={\bf F}{\bf y}+{\bf e}$ where the quantization distortion ${\bf e}$ is uncorrelated with ${\bf y}$ by construction. However, they are not independent. Hence, potentially better methods can be developed to exploit this dependency compared to the receivers presented in this section, which treat the quantization distortion simply as an independent colored noise. This is the main reason which motivates us to search for alternative solutions in Section 4. The Bussgang matrix ${\bf F}$ is given by \cite{one_bit_massive_mimo}
\begin{align}
{\bf F}=\sqrt{\frac{2}{\pi}}\text{diag}\big({\bf C}_{yy}\big)^{-1/2},
\end{align}
where ${\bf C}_{yy}=\mathbb{E}_{|{\bf \tilde{G}}}\{{\bf y}{\bf y}^H\}={\bf \tilde{G}}{\bf \tilde{G}}^H+{\bf \Sigma}$. $\mathbb{E}_{|{\bf \tilde{G}}}\{.\}$ and $\text{diag}(.)$ denote the conditional expectation given ${\bf \tilde{G}}$ and the diagonal part of a matrix, respectively. Here, ${\bf F}{\bf \tilde{G}}$ is the effective channel since it is the matrix multiplied with the data signal ${\bf s}$ in ${\bf \tilde{r}}={\bf F}{\bf \tilde{G}}{\bf s}+{\bf F}\bm{\mu}+{\bf e}$. The Bussgang-decomposition based quantization-aware MRC (BMRC), ZF (BZF), and MMSE (BMMSE) receivers are given by \cite{one_bit_receivers}
\begin{align}
&{\bf W}_{\text{BMRC}}={\bf \tilde{G}}^H{\bf F}^H, \label{eq:qa_mrc} \\
& {\bf W}_{\text{BZF}}=({\bf \tilde{G}}^H{\bf F}^H{\bf F}{\bf \tilde{G}})^{-1}{\bf \tilde{G}}^H{\bf F}^H, \label{eq:qa_zf} \\
& {\bf W}_{\text{BMMSE}}={\bf \tilde{G}}^H{\bf F}^H{\bf C}^{-1}_{\tilde{r}\tilde{r}} \label{eq:qa_mmse},
\end{align}
where ${\bf C}_{\tilde{r}\tilde{r}}=\mathbb{E}_{|{\bf \tilde{G}}}\{{\bf \tilde{r}}{\bf\tilde{r}}^H\}$ is given by \cite{one_bit_massive_mimo}
\begin{align}
&{\bf C}_{\tilde{r}\tilde{r}}=\frac{2}{\pi}\arcsin\bigg(\text{diag}\big({\bf C}_{yy}\big)^{-1/2}\Re\{{\bf C}_{yy}\}\text{diag}\big({\bf C}_{yy}\big)^{-1/2}\bigg)\nonumber \\
&+j\frac{2}{\pi}\arcsin\bigg(\text{diag}\big({\bf C}_{yy}\big)^{-1/2}\Im\{{\bf C}_{yy}\}\text{diag}\big({\bf C}_{yy}\big)^{-1/2}\bigg).
\end{align}
\section{ADMM-BASED SOLUTION}
Note that, the elements of the one-bit quantized signal, ${\bf r}=\text{sgn}({\bf z})$ are binary random variables which are obtained by scaling and shifting Bernoulli random variables. In the literature, there are some works \cite{deep_one_bit,near_ml_detector} that use the Bernoulli distribution in white noise corrupted models. In these works, the Q-function is effectively used since only the marginal distribution of $\{r_m\}$ is needed to develop quantization-aware algorithms. However, in our scenario where the binary random variables are dependent due to the effective colored noise, we cannot derive the joint distribution of  $\{r_m\}$ using the models in \cite{deep_one_bit,near_ml_detector}. In this paper, we will instead use the idea in \cite{signal_recovery} to express the signal detection as an optimization problem. Note that in \cite{signal_recovery}, there is only one parameter to be estimated and the resulting convex problem is solved using standard numerical methods which do not take the specific problem structure into account. In this paper, we will exploit the additional structure of the data vector ${\bf x}$ in \eqref{eq:z-x} and generalize the optimization problem in \cite{signal_recovery} for MIMO signal detection with hardware impairments by imposing more constraints. Then, an efficient formulation is proposed for the ADMM algorithm in order to obtain closed-form update equations. 

Note that if we were given the unquantized signal ${\bf z}$, the maximum-likelihood estimator of ${\bf x}$ without specifying any constraint on its structure is given as
\begin{align} \label{eq:ml_x}
{\bf \hat{x}}=({\bf H}^T{\bf C}^{-1}{\bf H})^{-1}{\bf H}^T{\bf C}^{-1}{\bf z},
\end{align}
which is obtained by minimizing the quadratic function $Q({\bf x},{\bf z})=({\bf z}-{\bf H}{\bf x})^T{\bf C}^{-1}({\bf z}-{\bf H}{\bf x})$ with respect to ${\bf x}$. If we insert the estimate of ${\bf x}$ in \eqref{eq:ml_x} into this quadratic function, we obtain a cost function to be minimized over ${\bf z}$ only which is
\begin{align} \label{eq:cost}
{\bf z}^T({\bf I}_{2M}-{\bf H}{\bf A})^T{\bf C}^{-1}({\bf I}_{2M}-{\bf H}{\bf A}){\bf z},
\end{align}
where ${\bf A}\in \mathbb{R}^{2K\times 2M}$ is defined as ${\bf A}\triangleq({\bf H}^T{\bf C}^{-1}{\bf H})^{-1}{\bf H}^T{\bf C}^{-1}$. If we use the required constraints that the elements of unquantized and quantized signals which are ${\bf z}$ and ${\bf r}$, have the same sign due to one-bit quantization, we obtain 
\begin{align}\label{eq:constraint1}
r_mz_m\geq 0, \ \ \ m=1,\ldots,2M.
\end{align}
In order to increase the signal detection performance, we can impose some additional constraints on ${\bf x}$. In this paper, we will focus on QPSK signaling for the data signals whose bit error rate (BER) performance does not degrade as other higher order constellation schemes due to the adverse effect of one-bit quantization \cite{ber_floor}. In this case, the elements of ${\bf x}$ are either $1/\sqrt{2}$ or $-1/\sqrt{2}$. Hence, the constraint on the elements of ${\bf x}$ becomes $x_k^2=1/2$, for $k=1,\ldots,2K$.

After defining the sign-refined vector ${\bf \tilde{z}}\triangleq \text{diag}({\bf r}){\bf z} \in \mathbb{R}_{\geq0}^{2M}$ and the local copy of ${\bf \tilde{z}}$, i.e., ${\bf t}={\bf \tilde{z}}$, in order to obtain closed-form updates, we obtain the quadratically-constrained quadratic programming (QCQP) problem
\begin{align}
& \underset{{\bf \tilde{z}},{\bf t},{\bf x}}{\text{minimize}} \ \ {\bf \tilde{z}}^T{\bf \tilde{B}}{\bf \tilde{z}} \label{eq:problem_cost} \\
& \text{subject to} \ \ \ {\bf t}={\bf \tilde{z}}, \ \ \ \ {\bf x}={\bf \tilde{A}}{\bf \tilde{z}}, \label{eq:problem_constraint1} \\
& \hspace{1.43cm} t_m\geq 0, \ \ \ m=1,\ldots,2M,  \label{eq:problem_constraint2} \\
& \hspace{1.43cm}  x_k^2=1/2, \ \ \   k=1,\ldots,2K \label{eq:problem_constraint3},
\end{align}
where ${\bf \tilde{A}}\triangleq{\bf A}\text{diag}({\bf r})$ and ${\bf \tilde{B}}\triangleq(\text{diag}({\bf r})-{\bf H}{\bf \tilde{A}})^T{\bf C}^{-1}(\text{diag}({\bf r})-{\bf H}{\bf \tilde{A}})$ for ease of notation. Even though we obtain a non-convex problem in this case, an efficient ADMM algorithm can be developed with closed-form updates. ADMM is a powerful first-order optimization method and has been proved to be effective in solving some non-convex problems \cite{consensus_admm, ozlem_improved_admm, nonconvex_admm1, nonconvex_admm2, nonconvex_admm3}. The problem we propose is a non-convex QCQP problem and there is a weak convergence result for these problems in \cite{consensus_admm} which states under some conditions, the ADMM iterations converge to a KKT point. Stronger versions of convergence are difficult to derive and it is left as future work.

In the following part, we will present the ADMM updates and derive their closed-form expressions.

\subsection{Closed-Form ADMM Updates}
The steps of the ADMM algorithm in scaled form \cite{boyd_admm} at the $i^{\textrm{th}}$ iteration are given as follows:
\begin{flalign}
& {\bf \tilde{z}}^{(i+1)} \leftarrow \argmin_{{\bf \tilde{z}}} \ \ {\bf \tilde{z}}^T{\bf \tilde{B}}{\bf \tilde{z}}+\rho||{\bf t}^{(i)}-{\bf \tilde{z}}+{\bf u}_1^{(i)}||^2 \nonumber \\
&\hspace{2.2cm}+\rho||{\bf x}^{(i)}-{\bf \tilde{A}}{\bf \tilde{z}}+{\bf u}_2^{(i)}||^2 \label{eq:admm1}, \\
& {\bf t}^{(i+1)} \leftarrow \argmin_{{\bf t}} \ \ \rho||{\bf t}-{\bf \tilde{z}}^{(i+1)}+{\bf u}_1^{(i)}||^2 \nonumber\\
& \hspace{2cm} \text{subject to} \ \ \  t_m\geq 0, \ \ m=1,\ldots,2M, \label{eq:admm2} \\
&{\bf x}^{(i+1)}\leftarrow \argmin_{{\bf x}} \ \ \rho||{\bf x}-{\bf \tilde{A}}{\bf \tilde{z}}^{(i+1)}+{\bf u}_2^{(i)}||^2  \nonumber \\
& \hspace{2cm} \text{subject to} \ \ \  x_k^2=1/2, \ \   k=1,\ldots,2K, \label{eq:admm3} \\
& {\bf u}_1^{(i+1)} \leftarrow {\bf u}_1^{(i)}+{\bf t}^{(i+1)}-{\bf \tilde{z}}^{(i+1)}, \label{eq:admm4} \\
& {\bf u}_2^{(i+1)}\leftarrow {\bf u}_2^{(i)}+{\bf x}^{(i+1)}-{\bf \tilde{A}}{\bf \tilde{z}}^{(i+1)}, \label{eq:admm5} 
\end{flalign}
where ${\bf u}_1 \in \mathbb{R}^{2M}$ and ${\bf u}_2 \in \mathbb{R}^{2K}$ are the scaled dual vector variables corresponding to the two equality constraints in \eqref{eq:problem_constraint1}, and $\rho>0$ is the penalty parameter used in the augmented Lagrangian \cite{boyd_admm}. The closed-form solutions of the problems in (\ref{eq:admm1})-(\ref{eq:admm3}) are derived as follows:
\begin{flalign}
& {\bf \tilde{z}}^{(i+1)}=\big({\bf \tilde{B}}/\rho+{\bf I}_{2M}+{\bf \tilde{A}}^T{\bf \tilde{A}}\big)^{-1} \times \nonumber \\
& \hspace{2.4cm} \big({\bf t}^{(i)}+{\bf u}_1^{(i)}+{\bf \tilde{A}}^T({\bf x}^{(i)}+{\bf u}_2^{(i)})\big)\label{eq:admm_1}, \\
& {\bf t}^{(i+1)}=\max{\big({\bf 0}_{2M}, {\bf \tilde{z}}^{(i+1)}-{\bf u}_1^{(i)}\big)}, \label{eq:admm_2} \\
& {\bf x}^{(i+1)}=\frac{1}{\sqrt{2}}\text{sgn}\big({\bf \tilde{A}}{\bf \tilde{z}}^{(i+1)}-{\bf u}_2^{(i)}\big), \label{eq:admm_3} 
\end{flalign}
where $\max(.,.)$ in \eqref{eq:admm_2} is performed element wise.

Although the projections in \eqref{eq:admm_2} and \eqref{eq:admm_3} are unique and optimum, the functions $\max(0,.)$ and $\text{sgn}(.)$ are not continuously differentiable at each point. Furthermore, it has been observed empirically that the harsh clipping effect of these functions did not bring any advantage compared to Bussgang-based quantization-aware MMSE receiver which performs the best among linear receivers in \cite{one_bit_receivers}. We believe that this is due to the harsh diminishing effect of other primal and dual optimization variables. One heuristic solution to preserve the effect of other variables' updates is to soften the projectors using soft approximations of $\max(0,.)$ and $\text{sgn}(.)$ which are softplus, $\text{softplus}(x)=\ln(1+e^x)$ and the hyperbolic tangent, $\tanh(x)$. Then, the modified updates in (\ref{eq:admm_2})-(\ref{eq:admm_3}) are given by
\begin{align}
&{\bf t}^{(i+1)}=\ln\big(1+e^{({\bf \tilde{z}}^{(i+1)}-{\bf u}_{1}^{(i)})}\big),  \label{eq:admm_modified_2} \\
& {\bf x}^{(i+1)}=\frac{1}{\sqrt{2}}\tanh\big({\bf \tilde{A}}{\bf \tilde{z}}^{(i+1)}-{\bf u}_2^{(i)}\big), \label{eq:admm_modified_3}
\end{align}
where all the operations are performed element wise. This modified version is shown to give good performance in the next section.

Note that the order of computational complexity of the ZF and MMSE-type receivers, and the proposed method is mainly determined by matrix inversions and multiplications. By some arrangement of sign refined matrices, it can be shown that the matrix inversion in \eqref{eq:admm_1} can be implemented only once per coherence block similar to the linear receivers. The number of matrix multiplications are determined by the number of iterations for the ADMM-based algorithm. It is shown in Section 5 that a substantially better performance is achieved by the proposed method compared to the linear receivers even when a small number of iterations is used. 
		\vspace{-0.3cm}
\section{Numerical Results}
		\vspace{-0.3cm}
In this section, we compare the BER performance of the proposed ADMM-based detection algorithm with the quantization-unaware and -aware linear receivers presented in Section 3 for the QPSK modulation scheme. Note that the performance of MMSE is very close to the ZF and BMMSE slightly outperforms BZF. Hence, we did not include the results of (B)ZF in the figures to simplify the presentation. The detection for the linear receivers is made based on the minimum Euclidean distance criteria between the processed signal and four possible constellation points. The iteration number and the penalty parameter for the proposed ADMM algorithm are set to 100 and $\rho=0.2$, respectively. The dual variables $\{{\bf u}_1,{\bf u}_2\}$ and the primal variables $\{{\bf t},{\bf x}\}$ are initialized as zero and standard Gaussian vectors, respectively. Since the sign of the elements of ${\bf t}$ are non-negative, we set ${\bf t}$ to the absolute value of it. The detection is made based on the sign of the elements of the vector ${\bf x}$ when the algorithm terminates. The simulation setup is based on \cite{kamil}. The user channels are modeled using independent  Rayleigh fading and $K$ users are distributed uniformly in a cell area of 250 m$\times$250 m. The path loss is calculated as $130+37.6\log_{10}(d)$ where $d$ is the distance between the user and BS in kilometers. The noise variance is $\sigma^2=2\times10^{-13}$. The heuristic uplink power control scheme in \cite{emil_book} is applied with maximum transmission power $p_{\text{max}}=0.1$ W and $\Delta=15$ dB. Hence, the transmission power of the $k^{\textrm{th}}$ user is reduced until its SNR becomes at most 15 dB above the worst user's SNR. The hardware quality coefficient for the BS and each UE is 0.98. In the following experiments, the BER is calculated based on 5000 different channel setups with 200 channel uses for each setup. 

In Fig.~1, we consider a scenario where the BS is equipped with $M=100$ antennas and there are $K=12$ users. Since the SNR of each user is different, we plot the average BER of the considered detectors with respect to the user index in ascending order of SNRs. Hence, as the user index increases from 1 to 12, the BER decreases due to the increased average SNR. Note that, as in accordance with the previous work \cite{one_bit_receivers}, the quantization-aware receivers, BMRC, BZF, and BMMSE, outperform their quantization-unaware counterpart receivers. However, the BER reduction is not significant compared to the huge performance gain achieved by the proposed ADMM-based algorithm. In fact, the average BER improvement is about two-fold for the worst SNR user and the BER improvement increases with SNR reaching approximately 14-fold  at user index 10 compared to the BMMSE receiver.

\begin{figure}[t!]
	\includegraphics[trim={2.9cm 0.1cm 0.2cm 1cm},clip,width=3.7in]{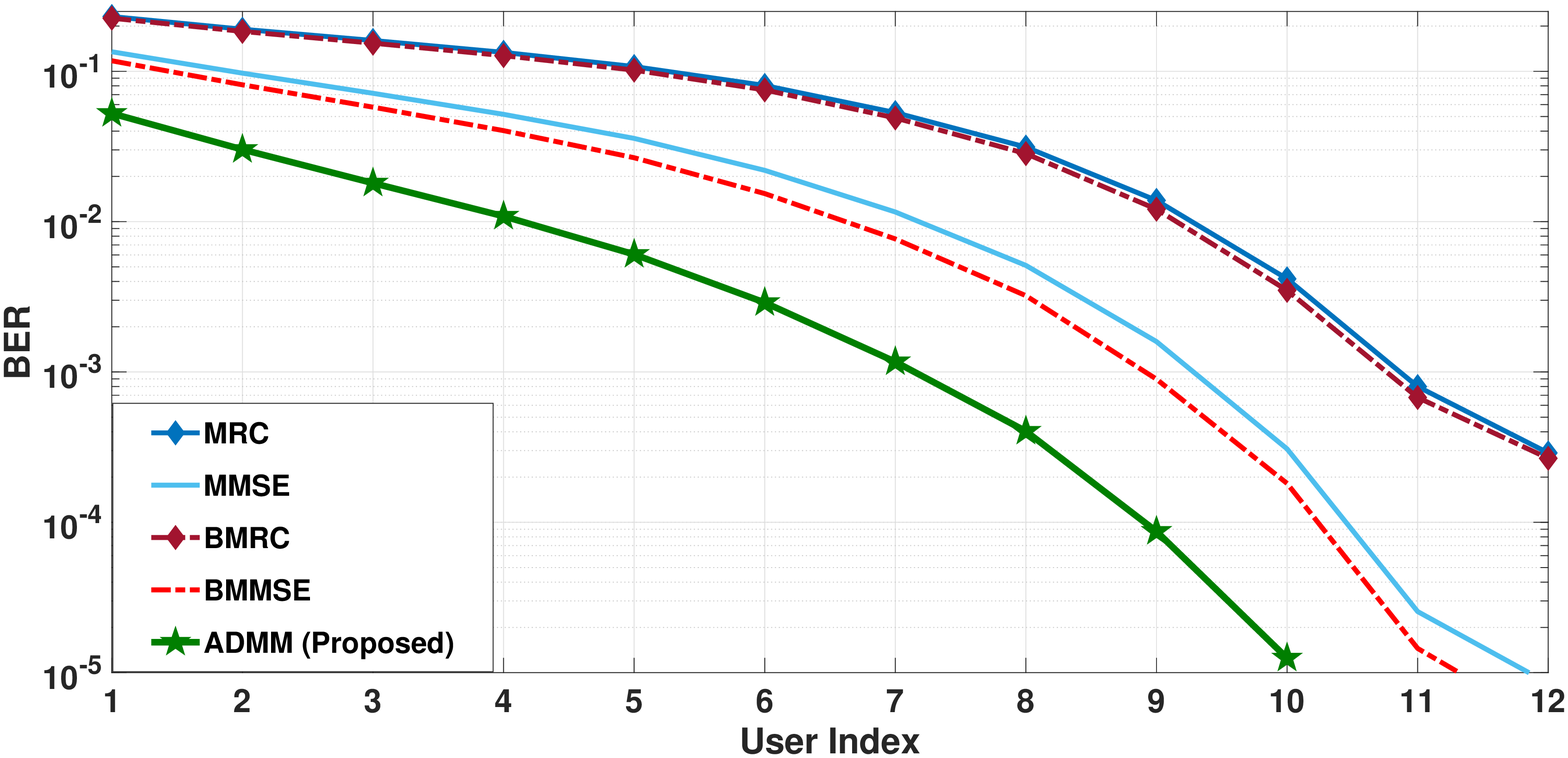}
	\vspace{-0.8cm}
	\caption{Average BER of the proposed ADMM-based algorithm and linear receivers versus user index in ascending order of SNRs for $M=100$, $K=12$.}
	\label{fig:fig1}
		\vspace{-0.3cm}
\end{figure}

In Fig.~2, we double the number of antennas and users to compare the performance of the proposed ADMM-based algorithm in a larger-size scenario. In this setup, $M=200$ and $K=24$, and we plot the average BER with respect to the user index. Similar to Fig.~1, the BER decreases with the user index, and hence SNR. The performance gap between the proposed algorithm and BMMSE is now larger compared to Fig.~1. There is a 30-fold BER improvement at user index 18, showing the effectiveness of the proposed algorithm for large number of BS antennas and users.
\begin{figure}[t!]
	\includegraphics[trim={2.9cm 0.1cm 0.2cm 1cm},clip,width=3.7in]{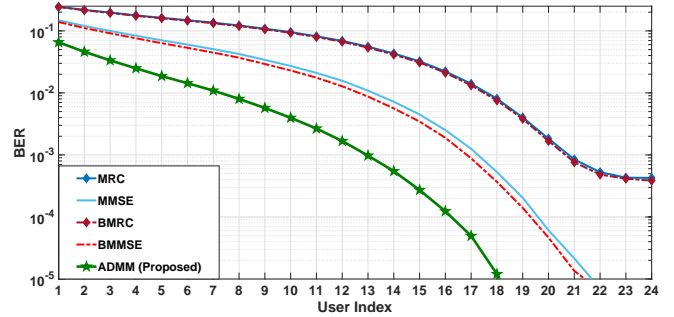}
		\vspace{-0.8cm}
	\caption{Average BER of the proposed ADMM-based algorithm and linear receivers versus user index in ascending order of SNRs for $M=200$, $K=24$.}
	\label{fig:fig2}
			\vspace{-0.3cm}
\end{figure}
		\vspace{-0.2cm}
\section{CONCLUSIONS AND FUTURE WORKS}
		\vspace{-0.2cm}
In this paper, we combine the joint effect of general additive hardware impairments and one-bit quantization in the uplink of a massive MIMO system. We have developed a new efficient ADMM-based algorithm with closed-form updates to improve the QPSK detection performance. A practical setup is considered with varying user SNRs in the numerical results. The new algorithm with softened projectors in its update equations outperforms the state-of-the-art linear receivers in the massive MIMO literature by exploiting the additional modulation and impairment characteristics. The proposed algorithm provides significant BER reduction for each user and the performance improvement increases with the user SNR and the number of BS antennas and users. 

We are planning to generalize the proposed algorithm to multi-bit quantization and higher-order modulations in future work. One other interesting research topic is to develop efficient algorithms in the presence of other time-varying hardware impairments. 
\vfill\pagebreak

%\bibliographystyle{IEEEbib}
%\bibliography{strings,refs}

\begin{thebibliography}{00}

	\bibitem{massive_mimo_reality} E. Bj\"ornson, L. Sanguinetti, H. Wymeersch, J. Hoydis, and T. L. Marzetta, \textquotedblleft Massive MIMO is a reality--What is next?: Five promising research directions for antenna arrays,\textquotedblright\ \emph{Digital Signal Processing,} vol. 94,
	pp. 3--20, Nov. 2019.


	\bibitem{one_bit_massive_mimo} Y. Li, C. Tao, G. S.-Granados, A. Mezghani, A. L. Swindlehurst, and L. Liu, \textquotedblleft Channel estimation and performance analysis of one-bit massive MIMO systems,\textquotedblright\ \emph{IEEE Trans.  Signal Process.,} vol. 65, no. 15, pp. 4075--4089, Aug. 2017.

	
		\bibitem{near_ml_detector} J. Choi, J. Mo and R. W. Heath, \textquotedblleft Near maximum-likelihood detector and channel estimator for uplink multiuser massive MIMO systems with one-bit ADCs,\textquotedblright in \emph{IEEE Trans. Commun.}, vol. 64, no. 5, pp. 2005--2018, May 2016.

		\bibitem{emil_nonideal} E. Bj\"ornson, J. Hoydis, M. Kountouris, and M. Debbah, \textquotedblleft Massive MIMO systems with non-ideal hardware: Energy efficiency, estimation, and capacity limits,\textquotedblright\ \emph{\  IEEE Trans. Inf. Theory,} vol. 60, no. 11, pp. 7112--7139, 2014.

	
	
	\bibitem{emil_book} E. Bj\"ornson, J. Hoydis, and L. Sanguinetti, \textquotedblleft Massive MIMO networks: Spectral, energy, and hardware efficiency,\textquotedblright\ \emph{Found. Trends Signal Process.,} vol. 11, no. 3-4, pp. 154--655, 2017.



	\bibitem{bruno_hardware} A. Papazafeiropoulos, B. Clerckx, and T. Ratnarajah, \textquotedblleft Rate-splitting to mitigate residual transceiver hardware impairments in massive MIMO systems,\textquotedblright\ \emph{IEEE Trans. Vehic. Tech.,} vol. 66, no. 9, pp. 8196--8211, Sept. 2017.

	
	\bibitem{emil_hardware} E. Bj\"ornson, L. Sanguinetti, and J. Hoydis, \textquotedblleft Hardware distortion correlation has negligible impact on UL massive MIMO spectral efficiency,\textquotedblright\ \emph{\  IEEE Trans. Commun.,} vol. 67, no. 2, pp. 1085--1098, Feb. 2019.

\bibitem{ozlem_iswcs} \"O. T. Demir and E. Bj\"ornson, \textquotedblleft Channel estimation under hardware impairments: Bayesian methods versus deep learning,\textquotedblright\ presented at \emph{ Int. Sympos. Wireless Commun. Systems (ISWCS),} Oulu, Finland, Aug. 2019.

	
	
		\bibitem{one_bit_massive_mimo2} F. Wang, J. Fang, H. Li, Z. Chen, and S. Li, \textquotedblleft One-bit quantization design and channel estimation for massive MIMO systems,\textquotedblright\ \emph{IEEE Trans. Vehic. Tech.,} vol. 67, no. 11, pp. 10921--10934, Nov. 2018.

			\bibitem{one_bit_receivers} L. V. Nguyen and D. H. N. Nguyen, \textquotedblleft Linear receivers for massive MIMO systems with
		one-bit ADCs,\textquotedblright\ unpublished paper, 2019. [Online]. Available: https://arxiv.org/abs/1907.06664

		\bibitem{ber_floor} A. Azizzadeh, R. Mohammadkhani, S. V. A.-D. Makki, and E. Bj\"ornson, \textquotedblleft BER performance analysis of coarsely quantized uplink massive MIMO,\textquotedblright\ \emph{Signal Processing}, vol. 161, pp. 259--267, 2019.

		
\bibitem{signal_recovery} S. Khobahi and M. Soltanalian, \textquotedblleft Signal recovery from 1-bit quantized noisy samples via adaptive thresholding,\textquotedblright\ in \emph{ 52nd Asilomar Conference on Signals, Systems, and Computers}, Pacific Grove, CA, USA, 2018, pp. 1757--1761.

	\bibitem{deep_one_bit} S. Khobahi, N. Naimipour, M. Soltanalian, and Y. C. Eldar, \textquotedblleft Deep signal recovery with one-bit quantization,\textquotedblright\ in \emph{ IEEE International Conference on Acoustics, Speech and Signal Processing (ICASSP)}, Brighton, United Kingdom, 2019, pp. 2987--2991.

\bibitem{one_bit_new1} Z. Cvetkovic, I. Daubechies, and B. F. Logan, \textquotedblleft Single-bit oversampled A/D conversion with exponential accuracy in the bit rate,\textquotedblright\ \emph{ IEEE Trans. Inf. Theory,} vol. 53, no. 11, pp. 3979--3989, Nov. 2007.



\bibitem{one_bit_new2} E. Masry and P. Ishwar, \textquotedblleft Field estimation from randomly located binary noisy sensors,\textquotedblright\ \emph{IEEE Trans. Inf. Theory,} vol. 55, no. 11, pp. 5197--5210, Nov. 2009.



\bibitem{one_bit_new3} A. Kumar, P. Ishwar, and K. Ramchandran, \textquotedblleft Dithered A/D conversion of smooth non-bandlimited signals,\textquotedblright\ \emph{IEEE Trans. Signal Process.,} vol. 58, no. 5, pp. 2654--2666, May 2010.



\bibitem{one_bit_new4} L. Jacques, J. N. Laska, P. T. Boufounos, and R. G. Baraniuk, \textquotedblleft Robust 1-bit compressive sensing via binary stable embeddings of sparse vectors,\textquotedblright\ \emph{IEEE Trans. Inf. Theory,} vol. 59, no. 4, pp. 2082--2102, Apr. 2013.


\bibitem{boyd_admm} S. Boyd, N. Parikh, E. Chu, B. Peleato, and J. Eckstein, \textquotedblleft Distributed optimization and statistical learning via the alternating direction method of multipliers,\textquotedblright\ \emph{Found. Trends Mach. Learn.,} vol. 3, no. 1, pp. 1--122, Jan. 2011.

\bibitem{lasso-admm} A. Elgabli, A. Elghariani, A. O. Al-Abbasi, and M. Bell, \textquotedblright Two-stage LASSO ADMM signal detection algorithm for large scale MIMO,\textquotedblright\ in \emph{\ 51st Asilomar Conference on Signals, Systems, and Computers}, Pacific Grove, CA, 2017, pp. 1660--1664.




	\bibitem{erik_book} T. L. Marzetta, E. G. Larsson, H. Yang, and H. Q. Ngo, \emph{\ Fundamentals of Massive MIMO.} Cambridge: Cambridge University Press, 2016.



	


	

	
	
	\bibitem{consensus_admm} K. Huang and N. D. Sidiropoulos, \textquotedblleft Consensus-ADMM for general quadratically constrained quadratic programming,\textquotedblright\ \emph{IEEE Trans. Signal Process.,} vol. 64, no. 20, pp. 5297--5310, Oct. 2016.

	

	
	\bibitem{ozlem_improved_admm} \"O. T. Demir and T. E. Tuncer, \textquotedblleft Improved ADMM-based algorithms for multi-group multicasting in large-scale antenna systems with extension to hybrid beamforming,\textquotedblright\ \emph{Digital Signal Processing,} vol. 93, pp. 43--57, 2019.

	\bibitem{nonconvex_admm1} R. Chartrand and B. Wohlberg, \textquotedblleft A nonconvex ADMM algorithm for group sparsity with sparse groups,\textquotedblright\ in  \emph{IEEE Int. Conf. Acoust., Speech, Signal Process. (ICASSP),} Vancouver, BC, 2013, pp. 6009--6013.



	
	\bibitem{nonconvex_admm2} X. Shen, L. Chen, Y. Gu, and H. C. So, \textquotedblleft Square-root Lasso with nonconvex regularization: An ADMM approach,\textquotedblright\  \emph{IEEE Signal Process. Lett.}, vol. 23, no. 7, pp. 934--938, Jul. 2016.


		\bibitem{nonconvex_admm3} M. \'A. V\'azquez, A. Konar, L. Blanco, N. D. Sidiropoulos, and A. I. P.-Neira, \textquotedblleft Non-convex consensus ADMM for satellite precoder design,\textquotedblright\ in \emph{IEEE Int. Conf. Acoust., Speech, Signal Process. (ICASSP),} New Orleans, LA, 2017, pp. 6279--6283.
	
	

	\bibitem{kamil} K. Senel, E. Bj\"ornson, and E. G. Larsson, \textquotedblleft Joint transmit and circuit power minimization in massive MIMO with downlink SINR constraints: When to turn on massive MIMO?,\textquotedblright\ \emph{\ IEEE Trans.  Wireless Commun.,} vol. 18, no. 3, pp. 1834--1846, Mar. 2019.



\end{thebibliography}

\end{document}